\documentclass[12pt]{article}
\begin{document}
\pagestyle{empty}
\begin{center}

{\Large {\bf Foreword: ASL Special Issue on Computational Astrophysics}}

\bigskip
Guest Editor: Lucio Mayer, {\it ASL Associate Editor}

\smallskip{\it Institute for Theoretical Physics, University of Z\"urich, Winterthurestrasse 190, 8057 Z\"urich, Switzerland}

\end{center}

\vskip 1.5truecm

Computational astrophysics has undergone unprecedented development over the last decade.
It has driven major progress in all fields of theoretical astrophysics and cosmology,
from the understanding of how the largest bound cosmic structures, galaxies and clusters of galaxies,
form and evolve, to the formation process of stars and planets. It has also led to 
a deeper understanding of individual astrophysical processes, such as supernovae explosions, jets or
gravitational instabilities and magnetorotational instabilities in accretion disks, that play 
a major role in several active research areas of astrophysics. While parallel supercomputer 
technology has boosted these advances  by allowing orders of magnitude increase in spatial, mass 
and time resolution, major breakthroughs have often been the product 
of improved numerical algorithms. Both particle-based and adaptive grid-based techniques have evolved in their
design and parallel performance, while simultaneously physical processes that were previously
neglected or poorly represented, such as radiation physics or magnetohydrodynamics, have
been included. Phenomenological models for unresolved sub-grid processes, inevitable in many
situations, have also become more physically motivated. Along with these conceptual and 
technical improvements pertaining to the simulations themselves, the 
creativity of individuals in designing simulations aimed at seeking answers
to unsolved problems across has been, and remains, central.
Creativity in this field is also becoming increasingly collective. Indeed
computational astrophysics and cosmology has grown into a field of its own, and
a raising number of national and international teams has been established. Their size is 
constantly growing, reflecting the increasing size of the calculations 
and the lifting demand for human and computational resources. This is fostering, among
other things, thorough comparisons between different codes, an exercise which is important and
necessary for the credibility and the productivity of the field itself.

\bigskip

But there is more to it. In the last few years there has been an increasing interest in interdisciplinary 
discussions among fields of computational astrophysics that had previously proceeded in completely independent ways, 
such as galaxy and star formation, or star and planet formation. The growing number of international 
workshops and conferences covering two or more
sub-fields reflect this tendency. We believe that this is extremely beneficial to the
community,  and at least as important as code comparisons. It is also an inevitable consequence of the 
increasingly multi-scale character of numerical simulations. Certainly 
one cannot understand galaxy formation without understanding star formation, or the formation of planets
without understanding the initial conditions of protoplanetary disks determined by physical
processes within star forming regions. Yet, it is only at this time that advances in computational
power and parallel algorithms allow us to conceive calculations
capable of modeling simultaneously different processes occurring at very different
scales. 

\bigskip

The challenge ahead of us is daunting. As calculations become more `multi-scale and complex,
simply understanding their results can become a major issue. This cannot be tackled by simply 
designing better parallel analysis tools, rather it requires nurturing a generation of
computational astrophysicists capable of a global view and general knowledge of astrophysical problems 
and of their modeling. This will be a necessary complement to the existing tendency towards deep specialization.
The physics behind many of the major open problems 
of astrophysics is the same, it just operates in different regimes. The same 
can be said for modeling techniques and strategies applied to different problems within astrophysics
and cosmology.
It is with an holistic approach in mind, and to foster interactions between the different areas of
research, that we present the Special Issue on Computational Astrophysics for the Advanced Science
Letters (ASL). Being a new multi-disciplinary scientific journal, Advanced Science Letters
offers a great opportunity to nurture a global perspective  and interactions 
between sub-fields of Computational Astrophysics.  ASL will act as a forum for the presentation
and discussion of novel work that attempts to bridge different areas.
This Special Issue collects 9 reviews contributed by a long list of distinguished authors on 9 key topics 
of modern astrophysics and cosmology in which
simulations had a pivotal contribution during the last decade. It intentionally covers
only simulations in the newtonian regime, leaving room for a future Issue on astrophysical simulations
in the relativistic regime. From asteroid collisions to the formation
of stars, from the physics and origin of galaxy clusters to large-scale phenomena such as 
the reionization of the Universe, it is a showcase of state-of-the-art computational
work in many different key areas of research. Most importantly, by presenting the 9 reviews in the
same volume,  this Special Issue is intended to stimulate the interest on the many physical and modeling
connections between these different topics. A list of the 9 Invited Reviews, with arXiv identifiers, can be found below.


\vskip 1truecm

{\Large \bf {Table of Contents: Invited Reviews}}

\bigskip

\begin{itemize}

\item{}
Collisional and rotational disruption of asteroids - arXiv:0906.4366
\smallskip

{\it Kevin Walsh, Patrick Michel, and Derek Richardson}

\item{}
Dynamical models of terrestrial planet formation - arXiv:0906.4369

\smallskip

{\it Jonathan Lunine, David O'Brien, Sean Raymond, Alessandro Morbidelli, Thomas Quinn, Amara Graps}

\item{}
Numerical simulations of disc-planet interactions - arXiv:0906.4347

\smallskip

{\it Richard Nelson and Sijme-Jan Paardekooper}

\item{}
Numerical star formation studies: a status report - arXiv:0906.4452

{\it Ralf Klessen, Mark Krumholz and Fabian Heitsch}

\item{}
Massive black hole binaries in the cosmic landscape - arXiv:0906.4339

{\it Monica Colpi and Massimo Dotti}

\item{}
The structure and evolution of cold dark matter halos - arXiv:0906.4340

{\it Juerg Diemand and Ben Moore}

\item{}
Cosmological simulations of galaxy clusters - arXiv:0906.4370

{\it Stefano Borgani and Andrey Kravtsov}

\item{}
Structure formation in the early Universe - arXiv:0906.4372

{\it Naoki Yoshida}

\item{}
Computer simulations of cosmic reionization - arXiv:0906.4348

{\it Hy Trac and Nick Gnedin}

\end{itemize}

\end{document}